# Investigation of plasmonic enhanced solar photothermal effect of Au NR@PVDF micro/nano-film

Shenyi Ding[a,b], Jixiang Zhang*[a,b], Cui Liu[b], Nian Li[b], Shudong Zhang[b], Zhenyang Wang*[b], Min Xi*[b,c]

[a] School of Mechatronics & Vehicle Engineering, Chongqing Jiaotong University, Chongqing, 400074, P. R. China

[b] Institute of Solid State Physics and Key Laboratory of Photovoltaic and Energy Conservation Materials, Hefei Institutes of Physical Science, Chinese Academy of Sciences, Hefei, Anhui 230031, P. R. China

[c] The Key Laboratory Functional Molecular Solids Ministry of Education, Anhui Normal University, Wuhu, Anhui, 241002, P. R. China

Corresponding authors:

Jixiang Zhang – Email: zhangjx@cqjtu.edu.cn

Zhenyang Wang − Email: zywang@iim.ac.cn

Min Xi − Email: minxi@issp.ac.cn

**Abstract**

Gold nanospheres (Au NSs) and gold nanorods (Au NRs) are traditional noble metal plasmonic nanomaterials. Particularly, Au NRs with tunable longitudinal plasmon resonance from visible to the near infrared (NIR) range were suitable for high efficient photothermal applications due to extended light receiving range. In this work, we synthesized Au NRs and Au NSs of similar volume, and subsequently developed them into Au NR/PVDF and Au NS/PVDF nanofilm, both of which exhibited excellent solar photothermal performance evaluated by solar photothermal experiments. We found that Au NR/PVDF nanofilm showed higher solar photothermal performance than Au NS/PVDF nanofilm. Through detailed analysis, such as morphological characterization, optical measurement, and finite element method (FEM) modeling, we found that the plasmonic coupling effects inside the aggregated Au NRs nanoclusters contributed to the spectral blue-shifts and intensified photothermal performance. As compared to Au NS/PVDF nanofilms, Au NR/PVDF nanofilm exhibited higher efficient light-to-heat conversion rate, because of the extended light receiving range and high absorbance, as the result of strong plasmonic interactions inside nanoclusters, which was further validated by monochromatic laser photothermal experiments and FEM simulations. Our work proved that the Au NRs have huge potential for plasmonic solar photothermal applications, and are envisioned for novel plasmonic applications.

**Introduction**

Solar photothermal technology is a most straightforward method to obtain clean energy.[1–3] In general, noble metal nanoparticles like gold and silver, due to their high light-to-heat conversion efficiency,[4] are of interests in the photothermal applications,[5] such as: solar steam generation (SSG),[6,7] photoacoustic virus inactivation,[8,9] photothermal therapy,[10] thermal imaging,[11] etc. Noble metal nanoparticle performs photothermal conversion via its localized surface plasmon resonance (LSPR) effects.[12,13] When the irradiated light is in resonance to the nanoparticle's surface plasmon frequency, a portion of light energy will be absorbed, and then be transformed in the form of heat as the result of electron-electron and electron-phonon interactions.[14,15]

In solar radiation spectrum, the solar emission intensive band covers the range from 400 nm to 2.5 microns,[3,16] and the near- to mid-IR emissions take more than half of the total radiation energy.[17] Therefore, efforts have been made to extend the light receiving range of the solar photothermal material to optimize the solar energy utilization. Unlike semiconductor plasmonic nanomaterials with adjustable optical properties that usually have near- to mid-IR plasmon,[18–20] noble metal nanoparticle's dipolar plasmon resonance usually tunes its LSPR via its shape.[21–23] As the simplest case, Au NR has independent collective electron oscillations along its transverse or longitudinal directions, respectively.[24] Particularly, its longitudinal plasmon can be varied from ~700 nm in the visible range to 1.2 microns in the near-IR as its aspect ratio (length to width ratio) changed from 2.3 to 6.5.[25]

For complex plasmonic hybridization system (plasmonic molecules), the interactions between individual plasmonic elements (plasmonic atoms) attribute to spectral shifts in the far-field and enhancement in the near-field.[26] When two nanoparticles approaches with their gap distance much smaller than their diameters, their dipolar plasmon hybridized and reformed.[27] For example, when two gold nanospheres (Au NSs) approach within the spacing distance ($g$) much smaller than their diameter ($d$), their dipolar resonances hybridized and red-shifted, and the corresponding spectral shift ($\Delta\lambda$) can be described by the semi-empirical equation as third order polynomial $d^3$ or exponential functional fits as: $\frac{\Delta\lambda}{\lambda} = 0.14 exp(\frac{-g/d}{0.23})$.[28] Moreover, if the close-packed nanoparticles can be assembled into one-dimensional nanochain, the dipolar resonance along the longitudinal axis can be further tailored into the red with the increased number of nanoparticles or decreased gap distance, and the corresponding spectral shifts can be described in the form of semi-empirical equation as: $\frac{\Delta\lambda}{\lambda_0} = exp(-\frac{m}{x})$,[29,30] where $\Delta\lambda$ is the peak shift ratio defined above, $x$ is the number of chain periods within the chain, $m$ is the characteristic interaction length, and $\lambda_0$ is the asymptotic value for the peak shift. If the gold nanoparticles were assembled into aggregated clusters, the coupled plasmonic interactions inside clusters will lead to both enhanced near- and far-field. For example, the patterned Au nanoparticle cluster arrays provides ~6 times higher Raman signal of tested bacteria species than non-patterned nor aggregated control samples.[31] If additional $SiO_2$ nanospheres can be well mixed with Au nanoparticle clusters, the developed Au/$SiO_2$ photonic-plasmonic nanoarrays leads to another ~3 times higher local E-field enhancement factor.[32] Moreover, in the case of nanofilm composite, the formed coupled plasmonic clusters contribute to improved light receiving capability, broaden light receiving range and high efficient photothermal performance.[7,33] For example, Kyuyoung Bae et al developed one large-area flexible thin-film black gold membrane consisting of self-aggregated metallic nanowire bundle arrays, which reached an average absorption of 91% under light receiving range of 400 - 2500 nm.[34] Chen et al developed the Au NP/PBONF composite films exhibit a solar-vapor conversion efficiency of 83%.[35]

However, the plasmonic coupling of Au NRs is a complicated case, because of the formed asymmetric alignment with higher degree freedom. Studies have shown mismatched formations contribute to the symmetry breaking Au NR dimer that of variety of near- and far-field shifts of their optical properties.[36,37] Particularly, the head-to-head (HH) alignment and the shoulder-by-shoulder (SS) alignment are the two most typical cases to start with. For HH alignment, the overall longitudinal dipolar modes are red-shifted, and the "hot spot" region in the gap indicates the near-field coupling, which can be applied for field enhancement molecular sensing applications, such as: surface enhanced Raman scattering (SERS) spectroscopy,[38] surface enhanced infrared absorption (SEIRA) spectroscopy,[39] surface enhanced fluorescence emission spectroscopy,[40,41] interference optical sensor,[42] etc. While for SS alignment, despite the enhanced intensity of overall transverse mode, the antibonding dark mode can be generated given the condition of asymmetric excitation, which can be used for energy storage applications, such as: transformation optics,[43] electromagnetically induced transparency (EIT),[44] Fano resonance,[45] etc. In the cases of bottom-up assembled Au NR nanocomposite, if high surface tension can be generated by appropriate nanoparticle surfactant and polymer medium, instead being uniformly distributed as colloidal suspensions, the nanorods are aggregated into coupled clusters due to surface tension, which are also statistically preferably in SS alignment.[46,47]

Although the spectral shifts of self-assembled or patterned individual Au NR cluster of different azimuthal alignments have been investigated, the aggregated Au NR cluster's coupling effects in bottom-up assembled nanocomposite in terms of film's optical response, especially the photothermal performance, to the best of our knowledge, have not been investigated statistically. In this manuscript, we chemically synthesized Au NRs and Au NSs, and subsequently developed them into Au NR/PVDF and Au NS/PVDF nanofilms, respectively. The nanoclusters were formed inside nanofilm via self-assembly due to surface tension of nanoparticles. Then we evaluated plasmonic coupling effects of aggregated nanoclusters through morphological characterization, optical measurement, and finite element method (FEM) modeling. Particularly, we found that the formed varies styles of Au NR nanoclusters inside nanofilm contributed to the spectral blue-shifts and increased optical density of the nanofilm. Compare to Au NS/PVDF nanofilms, Au NR/PVDF nanofilm exhibited higher efficient light-to-heat conversion rate, because of the extended light receiving range and higher absorbance, as the result of plasmonic coupling effects of the nanoclusters. Furthermore, we performed the monochromatic laser photothermal experiments and FEM simulations to validate our assumption. Our work proved that the Au NRs have huge potential for plasmonic solar photothermal applications, and are promising for vast range of novel plasmonic applications, such as: nanomedicine, tumor treatment, bio imaging, energy storage, etc.

**Results and Discussion**

The synthesized Au NRs and Au NSs were examined with transmission electron microscopy (TEM, jem-2100F) for morphological characterization. **Figure 1a** and **1b** show the typical TEM images of Au NRs and Au NSs, respectively. And the statistical counts of their size distributions were summarized in **Figure S1a1**, **S1a2** and **S1b** after processing the images with Nano Measure software. The averaged dimensions of Au NR were 68.9 ± 15.0 nm in length, 8.1 ± 1.0 nm in width with the aspect ratio of ~8.0. And the averaged diameter of Au NS is 18.8 ± 6.0 nm. Based on these measured dimensions, each individual Au NR and Au NS nanoparticle was estimated of approximately the similar volume as 3468.32 nm$^3$ and 3451.46 nm$^3$, respectively.

We characterized the optical properties of the synthesized colloidal Au NR and Au NS with UV-Vis-NIR spectrometer. Indicated as the black line in **Figure S2**, Au NR showed the transverse band (~520 nm) in the visible and longitudinal band (~830 nm) in the near-infrared (NIR), respectively. Particularly, the NIR resonance peak showed ~4 times higher intensity than transverse band, which indicated the monodispersed colloidal Au NRs product of high quality,[48] and was also confirmed by the TEM images in **Figure 1a**. On the other hand, as the red line in **Figure S2**, the LSPR resonance peak of Au NS is around 529 nm, and the sharp resonance peak suggested the uniform size distribution of Au NSs. Additionally, the high absorbance in the UV range (300 nm to 400 nm) indicates the interband transitions of electron oscillations.

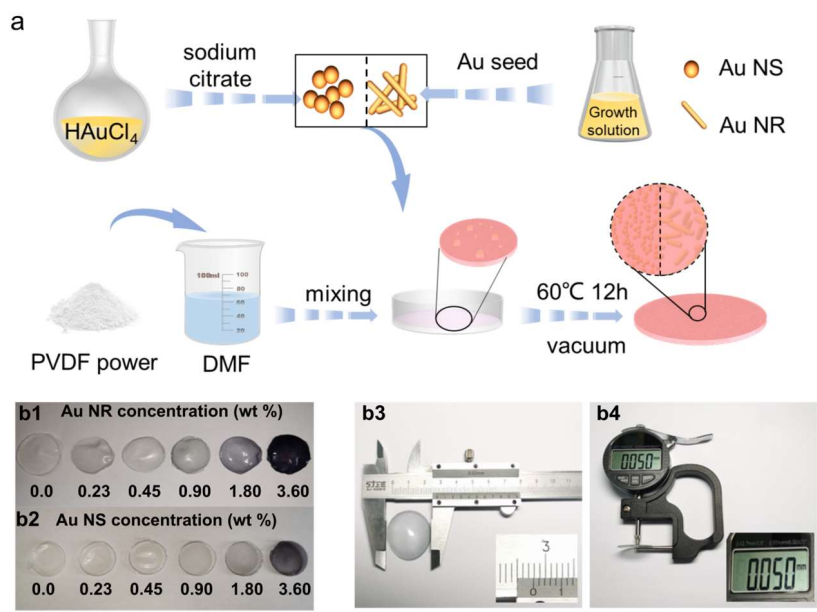

**Figure 1** Fabrication of Au NR/PVDF and Au NS/PVDF nanofilm. **(a)** Scheme depicts the nanofilm fabrication procedure; **(b1)** and **(b2)** Photographs of PVDF nanofilms with different Au loading concentrations; **(b3)** and **(b4)** Size measurement of nanofilm.

The synthesized gold nanoparticles were subsequently developed into photothermal nanofilms with polyvinyl difluoride (PVDF). We chose PVDF as the medium of photothermal film due to its abundance, convenience during its molding and polymerization, as well as high melting point (up to 170 °C) and stability after curing. Shown in **Figure 1c**, the nanocomposite was formed by mixing the synthesized gold nanoparticles with PVDF polymer solution under magnetic stirring, and then the nanofilm was formed by vacuum desiccation and oven baking (see **Methods** for experimental details). The resulting nanofilms were shown in **Figure 1d1** and **1d2**. For both Au NR/PVDF and Au NS/PVDF nanofilms, as increased weight concentration of loaded Au nanoparticles, the color of nanofilm becomes darker. Calibrated with ruler (**Figure 1d3** and **1d4**), the diameter of the nanofilm was determined as ~2.5 cm and the thickness was ~0.05 mm.

The optical characterization of nanofilms were performed via UV-Vis-NIR spectrometer measurement. The measured absorption spectrum were shown in **Figures 2a** and **2b**. Both of the Au NR/PVDF and Au NS/PVDF nanofilms showed intensified spectrum absorption with increased Au loading weight concentration. Particularly, since the spectrum of nanofilm were broad, especially for the nanofilms with small concentration of Au nanoparticles as relative low optical density, therefore, we determined the centered wavelength of these nanofilms via 2-steps fitting process. Shown in **Figure S3a – S3e,** we first smoothed the absorbance spectrum of nanofilms via polynomial fitting (shown in red line). Then, we fitted the absorption spectrum with multiple Gaussian peaks with natural life-time ($\sigma = 0.05$ of standard deviation) close to the peak width of monochromatic lasers. Then based on these fitted Gaussian peaks (shown in green), we leveled the background of the spectrum by choosing those peaks with dramatic intensities ($I > 0.6 I_{average}$), and rendered/described the rest of prominent values into an approximate peak that has the mean centered wavelength (shown in orange).

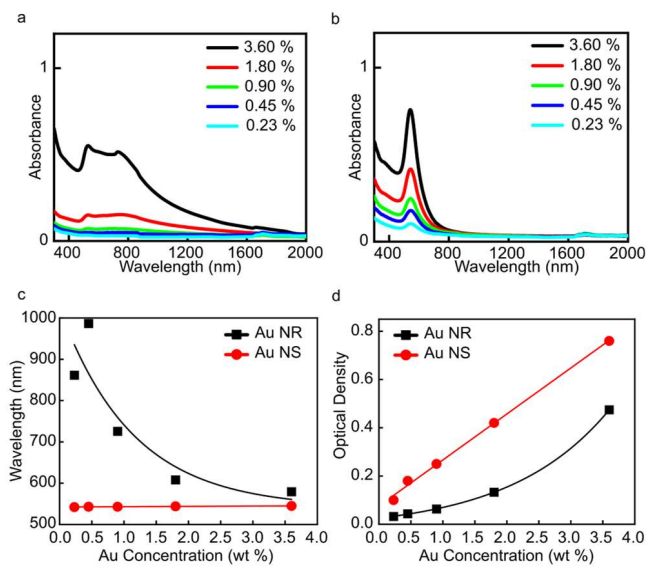

**Figure 2** Nanofilm optical characterization. **(a)** and **(b)** UV-Vis-NIR spectrum of Au NR/PVDF and Au NS/PVDF nanofilms with different Au loading concentrations; **(c)** Centered absorption wavelength of Au NR/PVDF (black line) and Au NS/PVDF nanofilms (red line) with different Au loading concentrations; **(d)** Optical density of Au NR/PVDF (black line) and Au NS/PVDF (red line) nanofilms with different Au loading concentrations.

Shown in **Figure 2c**, the Au NR/PVDF nanofilm showed spectral blue shifts from ~980 nm to ~580 nm as the Au loading concentration increased from 0.23 % to 3.60 %. While Au NS/PVDF nanofilm showed almost the same spectral peak with increased Au loading concentration. The nanofilms' optical densities of Au NR/PVDF and Au NS/PVDF nanofilms were summarized in **Figure 2d**. Both Au NR/PVDF and Au NS/PVDF nanofilms exhibited increased optical density (defined as plasmonic peak intensity) with increased Au loading concentration. Notably, Au NS/PVDF nanofilm showed higher optical density than Au NR/PVDF nanofilm, while Au NR/PVDF nanofilm had wider light receiving range than Au NS/PVDF nanofilm. In other words, the nonlinear relationship between the optical density of Au NR/PVDF nanofilm and the Au loading concentration, as well as the spectral blue-shifts of the Au NR/PVDF nanofilm, suggested the existence of the strong plasmonic coupling effects of formed Au NRs aggregations.

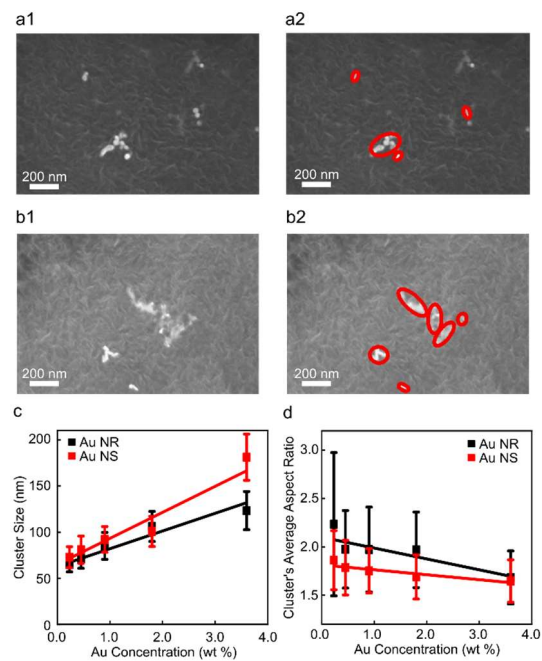

**Figure 3** Morphological characterization of nanofilms. **(a1)** and **(b1)** Representative SEM images to show Au NS/PVDF nanofilm, and the aggregated cluster marked by the programmed codes; **(a2)** and **(b2)** Representative SEM images to show Au NR/PVDF nanofilm, and the aggregated cluster marked by the programmed codes; **(c)** Averaged sizes of aggregated clusters inside Au NS/PVDF and Au NR/PVDF nanofilm as function of Au loading weight percentage, respectively; **(d)** Averaged aspect ratios of aggregated clusters inside Au NS/PVDF and Au NR/PVDF nanofilm as function of Au loading weight percentage.

In order to observe the agglomerated clusters of Au nanoparticles in the nanofilm, we characterized morphology of gold nanoparticles inside PVDF nanofilm. We examined the gold nanoparticles' distribution via characterization of the sliced sample nanofilm with scanning electronic microscopy

(SEM). The representative SEM images of Au NR/PVDF and Au NS/PVDF nanofilm were shown in **Figure 3a1** and **3b1**. Other SEM images with varied distributions of gold nanoparticles in the nanofilm were shown in **Figure S3** (Au NR/PVDF nanofilm) and **S4** (Au NS/PVDF nanofilm), respectively. Specifically, Au NRs in the nanofilms were aggregated into the clusters during the solvent evaporation in the process of nanofilm fabrication, and both of the number and size of agglomerations increased with Au loading weight concentration. Similarly, Au NSs formed fewer agglomerations in the medium of PVDF than Au NRs given the same Au loading concentration.

In order to statistically analyze the formed clusters in the nanofilms, we processed the characterized SEM images with programmed codes for computational vision (CV) recognition (see **Method** for coding details). **Figure 3a2** and **3b2** showed the processed images originated from **Figure 3a1** and **3b1.** The aggregated clusters were identified and approximated with elliptically fits. Marked in red, the size of cluster was averaged by the longitudinal axis and transverse axis of fitted curves. The statistical counted size distributions were summarized in **Figure 3c**. For both of Au NR/PVDF and Au NS/PVDF nanofilms, the size of formed clusters inside nanofilms increased linearly with the Au loading concentrations, which confirmed our assumption mentioned above that the size of clusters increased with Au loading concentration. It is intriguing that for Au NR/PVDF nanofilms, the size of the nanoclusters increased with Au NRs loading concentrations, which lead to the blue shifts of centered wavelength, which is different from the cases of Au NS/PVDF nanofilms that slightly red-shifts.[20,49] Therefore, we hypothesized that the decreased aspect ratio (that usually contributes to blue-shifts[23,50]) outweighs the size effects (that contributes to red-shifts) during the increased loading weight concentrations of Au NR/PVDF nanofilms.

Therefore, we summarized the aspect ratios of formed aggregated clusters in Au NR/PVDF and Au NS/PVDF nanofilms in **Figure 3d**, respectively. Specifically, the aspect ratio of formed clusters in Au NR/PVDF nanofilms decreased from ~2.3 to ~1.5, which suggested that the formed spherical clusters with resonance in the visible and high intensity that contribute to the spectral blue-shifts and increased optical density of Au NR/PVDF nanofilm in **Figure 2c** and **2d**, respectively.

We assumed that the increased size of formed agglomerated clusters inside nanofilm would lead to strong plasmonic coupling effects of clusters as well as improved photothermal performance, therefore, we evaluated the solar photothermal performance of Au NR/PVDF and Au NS/PVDF nanofilm to validate the assumption. Shown in **Figure 4a**, the nanofilms were illuminated by a simulated solar light source, and the temperature changes of the sample nanofilm were monitored by an infrared camera. **Figure 4b** shows the representative recorded thermal images of Au NR/PVDF and Au NS/PVDF nanofilms after receiving different time of solar illumination. Both of the nanofilms showed significant elevated temperature after ~300 s. Experimental measured solar photothermal performance of Au NR/PVDF and Au NS/PVDF nanofilm were summarized as function of time in **Figure 4c** and **4d**, respectively.

The measured solar photothermal heating rate and equilibrium temperature of Au NR/PVDF and Au NR/PVDF nanofilm were summarized as function of Au weight concentration in **Figure 4g** and **4h** as dashed curves, respectively. As the weight percentage increased in the cell solution from 0 % to 3.60 %, the warming rate of Au NS/PVDF (red dashed curves) increased from 0.07 °C/s to 0.12 °C/s, compared to that of Au NR/PVDF (black dashed curves) from 0.08 °C/s to 0.17 °C/s. Similarly, the temperature at equilibrium of Au NS/PVDF (red dashed curves) increased from 30 °C to 37 °C as the weight percentage in the nanofilm increased from 0 % to 3.60 %, compared to that of Au NR/PVDF (black dashed curves) increased from 30 °C to 40 °C.

To further validate the plasmonic photothermal performance of nanofilms, we applied numerical simulation to model the photothermal effects. Shown in **Figure S5**, the simulation was performed with 2D heat transfer module of COMSOL Multiphysics to predict the temperature changes of nanofilm under the solar illumination (see **Method** for simulation details). The predicted temperature changes of Au NR/PVDF and Au NS/PVDF nanofilm in **Figure 4e** and **4f** basically follow the trend of experimental measured results in **Figure 4c** and **4d**.

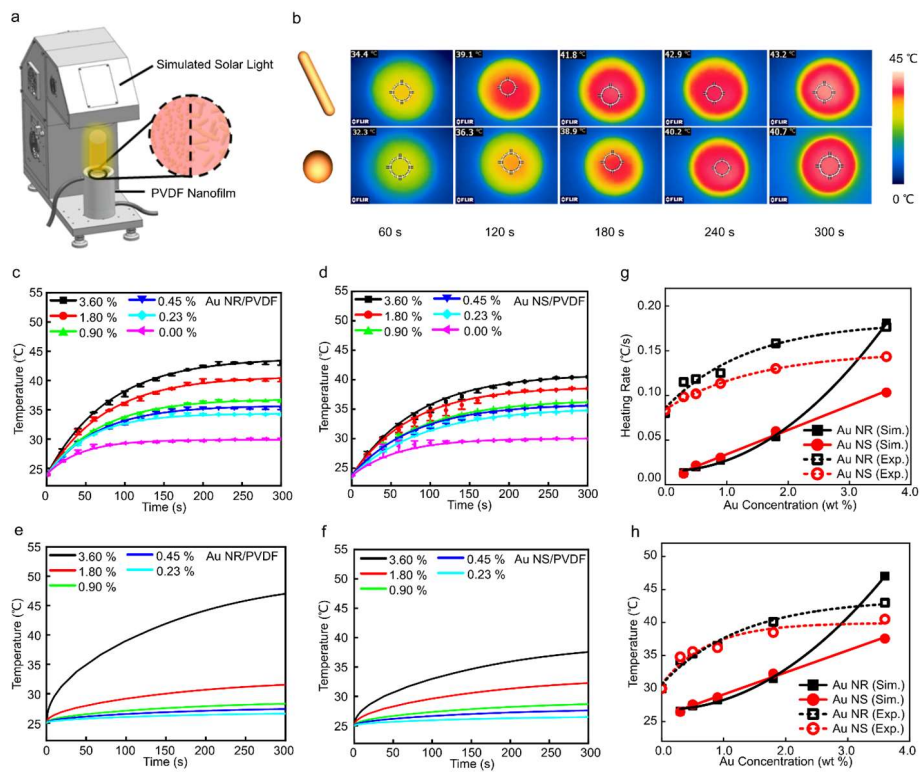

**Figure 4** Solar photothermal performance of nanofilms. **(a)** Scheme depicts the solar photothermal experimental setup; **(b)** Recorded thermal images of Au NR/PVDF (upper row) and Au NS/PVDF (lower row) nanofilm of different illumination time obtained by infrared camera; **(c)** and **(d)** Experimental measured solar photothermal performance of Au NR/PVDF and Au NS/PVDF nanofilm as function of time; **(e)** and **(f)** Simulated solar photothermal performance of Au NR/PVDF and Au NS/PVDF nanofilm as function of time; **(g)** Experimental and simulated solar photothermal heating rate of Au NR/PVDF and Au NR/PVDF nanofilm as function of loaded Au weight concentration; **(h)** Experimental and simulated equilibrium temperature of Au NR/PVDF and Au NR/PVDF nanofilm as function of loaded Au weight concentration.

The simulated solar photothermal heating rate and equilibrium temperature of Au NR/PVDF and Au NR/PVDF nanofilm were also summarized in solid lines as function of Au weight concentration in **Figure 4g** and **4h**, respectively. Both experimental and simulated results indicated that the Au NR/PVDF nanofilm outperformed Au NS/PVDF nanofilm in terms of photothermal heating rate and temperature at equilibrium due to the stronger plasmonic coupling effects and photothermal performance of Au NRs clusters in the nanofilm. It should be noticed that the simulated heating rate and temperature at equilibrium of Au NR/PVDF nanofilm exhibited nonlinear growth rate with increased Au loaded weight concentration, which differed from experimental measured data that both of heating rate and temperature at equilibrium's growth rate decreased and eventually saturated with increased Au loaded weight concentration. We attributed this deviation to the underestimation of heat dissipation rate in the simulation model.

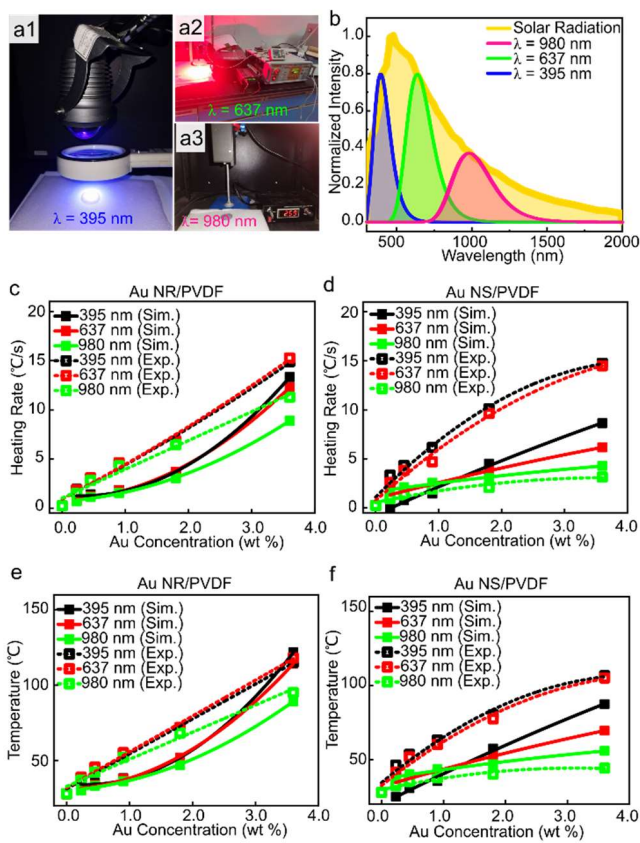

**Figure 5** Monochromatic laser photothermal performance of nanofilms. **(a)** Photograph of monochromatic laser with centered wavelength of 395 nm **(a1)**, 637 nm **(a2)** and 980 nm **(a3)**; **(b)** Solar spectrum and monochromatic laser spectrum; **(c)** and **(d)** Experimental measured and calculated heating rate of Au NR/PVDF and Au NS/PVDF nanofilm as function of time with illumination of different monochromatic laser as function of Au weight concentration; **(e)** and **(f)** Experimental measured and calculated equilibrium temperature of Au NR/PVDF and Au NS/PVDF nanofilm as function of time with illumination of different monochromatic laser as function of Au weight concentration.

We further quantified each band's contribution during the solar photothermal performance via evaluating the warming rate with illumination of monochromatic UV, visible and NIR laser. Shown in **Figure 5a1 – 5a3**, we performed laser photothermal warming by selecting monochromatic LED laser of UV ($\lambda$ = 395 nm), visible ($\lambda$ = 637 nm), and NIR ($\lambda$ = 980 nm) as the light sources. The experimental measured T-t curves of Au NR/PVDF and Au NS/PVDF nanofilm were summarized in

**Figure S6a – S6c** and **Figure S6d – S6f**, respectively. Furthermore, we validated the heat transfer situation of nanofilms under monochromatic laser illumination through numerical simulation. The calculation was performed by describing monochromatic laser light sources of different wavelengths, 395 nm, 637 nm and 980 nm approximately into Gaussian beams (**Figure 5b**). The simulation T-t curves of Au NR/PVDF and Au NS/PVDF nanofilm were summarized in **Figure S7a – S7c** and **Figure S7d – S7f**, which basically matched the experimental measured results in **Figure S6a – S6c** and **Figure S6d – S6f**.

The experimental measured and simulated warming rate of different monochromatic lasers were summarized as function of loaded Au weight concentration in **Figure 4c** (Au NR/PVDF) and **4d** (Au NS/PVDF), and the experimental measured and simulated equilibrium temperature of different monochromatic lasers were summarized as function of loaded Au weight concentration in **Figure 4e** (Au NR/PVDF) and **4f** (Au NS/PVDF), respectively. The simulated results were in excellent agreement with the experimental value. UV and visible monochromatic lasers outperformed NIR laser in both of the Au NR/PVDF and Au NS/PVDF nanofilms photothermal process in terms of heating rate and equilibrium temperature, which indicated that UV and visible bands contributed more than the NIR band in the process of solar photothermal performance. Notably, the Au NR/PVDF nanofilm exhibited much more significant higher heating rate and equilibrium temperature than Au NS/PVDF nanofilm, which demonstrated that the aggregated Au NR clusters in the Au NR/PVDF nanofilm contributed to the boosted solar photothermal performance by extending the light receiving range to the longer wavelength.

In order to further quantify the photothermal performance of Au NR/PVDF and Au NS/PVDF, we evaluate their light conversion efficiency, which was calculated by the energy balance function, and the light-to-heat conversion rate can be estimated as : $\eta = E_{in}/E_0$ , where $E_{in} = \sum m_i C_{p,i}(T_n - T_0)$ described the absorbed energy by the Au NR/PVDF or Au NS/PVDF nanofilms, and $E_0 = Pst$ described the overall energy output provided by the light source (see **Methods** section for detailed information). The calculated light-to-heat conversion efficiency of solar photothermal and monolaser photothermal were summarized in **Table S1** and **S2**, respectively.

Shown in **Figure S1**, as the loading weight of Au increased from 0.23 % to 3.60 %, both Au NR/PVDF and Au NS/PVDF nanofilms showed increased solar photothermal conversion efficiency from ~1.0 % to 12.0 %, though Au NR films showed slightly higher conversion efficiency than Au

NS films. **Table S2** showed Au NR films and Au NS films under different illumination of monolasers with wavelength of 395 nm, 637 nm and 980 nm. Au NR films showed ~10.0 % higher photothermal conversion rate than Au NS films at all cases of monolaser illumination. This result is consistent with our observations and assumptions mentioned above that the aggregated Au NR clusters in the Au NR/PVDF nanofilm contributed to the boosted solar photothermal performance by extending the light receiving range to the longer wavelength, and to some extent, it proved the Au NR/PVDF nanofilm's potential in utilizing safer and bio-friendly IR light in natural solar light.

**Conclusion**

In this manuscript, we chemically synthesized Au NRs, and developed them into Au NR/PVDF nanofilms. Through morphological characterization and statistical analysis, we found that formed varies styles of coupled Au NR clusters in the nanofilm, which contributed to the spectral blue-shifts and boosted photothermal performance. Particularly, through quantitative analysis of nanofilm's composition, optical characteristics, and photothermal response under solar irradiation and/or under monochromatic laser illumination, we found that the as-the synthesized Au NRs formed Au NR clusters, which significantly improved the solar energy conversion rate via extending the light harvest range and nanofilm's optical density. We believe our work have demonstrated that the Au NRs have huge potential for plasmonic solar photothermal applications, and are promising for vast range of applications, such as: nanomedicine, cancer therapy, tumor imaging, solar harvesting, etc.

**Methods**

*Chemicals and Materials*

Gold chloride hydrate, hydroquinone (AR), sodium borohydride (96 %), N,N-Dimethylformamide (DMF, AR, 96 %) were purchased from Aladdin. Hexadecyl trimethyl ammonium bromide (CTAB, AR), Trisodium citrate dehydrate (AR), Polyvinylidene fluoride (PVDF, MW = 1 million) were purchased form Sinopharm Chemical Reagent. Sodium hydroxide (AR), silver nitrate solution was purchased from Macklin. Ultrapure water (18.2 MΩ cm$^{-1}$) was produced by a Millipore water purification system.

*Au Nanoparticles Synthesis*

The synthesis method of Au NRs were applied as seed-mediated growth technique developed by Leonid Vigderman et al.[51] The Au seeds were first prepared by fast reducing Au precursor with sodium borohydride, and were subsequently developed into Au NRs in the bath solution. Typically, 0.01 M sodium borohydride was dissolved in 0.01 M sodium hydroxide solution. Under rapid stirring, 0.46 mL of freshly prepared sodium borohydride solution was added into the $HAuCl_4$ solution (10 mL, 0.5 mM) in 0.1 M CTAB. The color of the solution changed from yellow to light brown, and then was used for the seed solution.

For Au NR growth, 1 mL silver nitrate solution (0.1 M) was added into $HAuCl_4$ solution (100 mL, 0.5 mM) with 0.1 M CTAB, and then hydroquinone aqueous solution (5 mL, 0.1 M) was added. The solution was kept for stirring at 30 °C until completely clear, then 1.5 mL of seed solution was added. The color of the growth solution gradually changed from colorless to wine red. After aging overnight, the resulting Au NRs were finally obtained after being washed repeatedly by centrifugation and re-dispersed in water.

The synthesis of Au NSs was adopted from the Turkevich method,[52] which was to reduce the Au precursor with trisodium citrate under vigorously stirring and boiling. Typically, 50 mL $HAuCl_4$ solution (10 mM) is heated to boiling in an oil bath at 137 °C. Under vigorous stirring, 1.5 mL trisodium citrate aqueous solution (0.01 g/mL) was quickly added into the boiling $HAuCl_4$ solution. The color changes from light yellow to dark blue and then wine-red. The solution was kept refluxed for another 30 min, and then washed with deionized water for three times and finally re-dispersed in water.

*Au Nanoparticles/PVDF Nanofilms Synthesis*

PVDF powder and DMF solution were mixed in a beaker at a mass fraction ratio of 1:10, the beaker was sealed to prevent evaporation of DMF, and the mixture was magnetically stirred overnight to ensure complete dissolving of PVDF. After centrifugation of the gold nanoparticles, the pellets were re-dispersed in 100 μL DMF with subsequent ultrasonication. 1 mL of PVDF/DMF solution and 100 μL Au/DMF were placed in a glass Petri dish. The solution was mixed via stirring and excessive air bubbles were removed. The solution was placed in a vacuum drying oven at 60 °C to remove the DMF solvent to obtain dry Au/PVDF nanofilms. The resulting Au weight concentration in the nanofilm was quantitative determined via inductively coupled plasma-mass spectrometry (ICP-MS) study.

*Nanofilm Optical Measurement*

The optical responses of Au NR, Au NS colloidal solution, and Au/PVDF nanofilms were characterized by UV-Vis-NIR spectrometer UV-3600 (SHIMADZU). The sample spectrum was recorded at wavelength range of 400 − 1200 nm and slit width of 2 nm. And the recorded spectra was normalized and fitted by the Gaussian peak to determine the resonance peak and FWHM.

*Nanofilm Morphological Characterization*

The morphology of nanofilm characterization was performed with a field emission scanning electron microscope (FESEM, SU8020, HITACHI) with the acceleration voltage of 10 kV. The nanofilm was cut into thin slices with thickness of ~50 nm with ultrathin nanoblade slicer for SEM examination.

*Computational Vision Recognition*

The obtained SEM images were processed by programmed Python CV codes to identify formed clusters (codes were available on https://github.com/MinXi2021/clusters_recognition_CV_codes.git). Typically, the images were firstly leveled the threshold, and then the formed clusters were identified via approximating with polygon. The size of cluster was averaged by the longitudinal axis and transverse axis of fitted curves.

*Photothermal Experiments*

For solar photothermal experiment, Au/PVDF nanofilms with different concentrations of Au loaded weight concentrations were placed under a solar simulator (7IS0503A). The temperature of nanofilms were recorded by an infrared thermal imaging camera (FLIR C2) every 20 seconds with an infrared camera until the temperature change leveled off. The light intensity of the solar simulator was adjusted to a solar constant ($1367 \pm 7$ W/m$^2$) calibrated by an optical power meter.

For the study of monochromatic laser photothermal experiment, the LED laser with centered wavelength of 395 nm, 637 nm and 980 nm were applied, and the output irradiance power were adjusted to 1.8 W (with the spot of illuminated area of ~5 cm$^2$) calibrated by an optical power meter.

All measurements were performed for 3 times to check the reproducibility.

*Solar Photothermal Simulations.*

The heat transfer module of COMSOL Multiphysics was used to calculate photothermal response. Shown in **Figure S5**, the sample nanofilm (indicated in blue meshed region in the bottom) was modeled, which was described as cylinder with radius = 1.25 cm and thickness = 0.005 cm in the space of air. The input radiance was set perpendicular to the sample tube at Z direction.

The partial differential equations **Eq. 1** and **Eq. 2** were solved to calculate the temperature distribution of the system. **Eq. 1** describes the thermal equilibrium of the overall system, where $Q_r$ is the generated heat; $\rho_{air}$, $\rho_{PVDF}$ and $C_{air}$, $C_{PVDFn}$ are the density and heat capacity of air and solution, respectively.

$$\frac{\partial Q_r}{\partial t} = \rho_{air} C_{air} \frac{\partial T_{air}}{\partial t} + \rho_{PVDF} C_{PVDF} \frac{\partial T_{PVDF}}{\partial t} \qquad \textbf{Eq. 1}$$

**Eq. 2** describes the photothermal process, where $K_{PVDF}$ is the nanofilm's absorption coefficient, $G$ is the radiation power density.

$$\frac{\partial Q_r}{\partial t} = K_{PVDF} G \qquad \textbf{Eq. 2}$$

For the calculation of cell solution's heat transfer under solar light illumination, similar to our previous study,[19] we approximated and then processed the solar spectrum into a series of elemental beams, and each elemental beam was described by the Gaussian fit curve (**Eq. 3**) as:

$$I_i = \frac{I_0}{\sqrt{2\pi}\sigma} e^{-\frac{(f-f_0)^2}{2\sigma^2}} \qquad \textbf{Eq. 3}$$

where $I_i$ is each processed beam's intensity (the laser input power was set as 1300 W/m$^2$), $I_0$ is the corresponding intensity of solar spectrum, $f_0$ is the central frequency of the beam, and $\sigma$ is the standard deviation that is proportional to $D_i$ (beam diameter of $1/e^2$ width) defined as $\sigma = \frac{D_i}{f_0}$, and we chose $\sigma = 0.02$ for simulation.

Therefore, the solar power intensity can be described as:

$$G = \sum a_i \cdot I_i \qquad \text{Eq. 4}$$

Where $a_i$ is each approximated elemental Gaussian beam's portion of overall solar irradiance power, $I_i$ is the elemental beam's corresponding intensity to each frequency. And partial differential equation to be solved can be rewritten as **Eq. 5**:

$$\frac{\partial Q_r}{\partial t} = \sum \kappa_i \cdot a_i \cdot I_i = \rho_{air} C_{air} \frac{\partial T_{air}}{\partial t} + \rho_{PVDF} C_{PVDF} \frac{\partial T_{PVDF}}{\partial t} \qquad \text{Eq. 5}$$

Where $a_i$ can be calculated from the measured optical density of the nanofilm from **Eq. 6** as:

$$a_i = \sum \kappa_i \qquad \text{Eq. 6}$$

Where the absorption coefficient is described as the sum of $\kappa_i$, which is the optical density at each frequency.

The model was meshed into ~3000 elements and the simulation performed with PARSIDO time dependent solver to reach the relative tolerance of 0.01.

For the heat transfer simulation monochromatic laser illumination, monochromatic laser light sources with centered wavelength of 395 nm, 637 nm and 980 nm were approximately described as Gaussian beams. Since $\sigma$ of monochromatic laser beam is usually larger than 0.02, we choose $\sigma = 0.02$ to describe the laser intensity with Gaussian fit. The laser input power was set as 18000 W/m².

*Photothermal Conversion Efficiency Calculation*

The photothermal conversion efficiency under simulated solar irradiation can be obtained from the following equation:

$$\eta = \sum m_i C_{p,i} (T_n - T_0)/Pst \qquad \text{Eq. 7}$$

where $m_i$, $C_{p,i}$ are the mass ($m_{Au/PVDF} \sim 0.2g$) and heat capacity ($C_{p,Au} = 0.13$ J·g⁻¹·K⁻¹), $C_{p,PVDF} = 1.17$ J·g⁻¹·K⁻¹) of Au and PVDF, respectively, $T_n$ is the highest temperature, $T_0$ is the initial temperature, $P$ is the solar irradiation power (~ 1300 W/m²), $s$ is the film area (~ 6 cm²), and $t$ is the irradiation time.

The photothermal conversion efficiency of Au/PVDF films under laser irradiation was calculated by the energy balance function.

$$\sum m_i C_{p,i} \frac{dT}{dt} = Q_{in} - Q_{ext} \qquad \text{Eq. 8}$$

where $T$ is the temperature, $Q_{in}$ is the heat absorbed by the Au/PVDF film, and $Q_{ext}$ is the heat dissipated by the surroundings.

$$Q_{in} = I \cdot A_{abs} \cdot \eta \qquad \text{Eq. 9}$$

where $I$ is the laser power, $A_{abs}$ is the absorbance, and $\eta$ is the photothermal conversion efficiency.

$$Q_{ext} = hs(T_n - T_0) \qquad \text{Eq. 10}$$

where $h$ is the heat transfer coefficient.

By fitting the temperature vs. time curve after turning off the laser.

$$T(t) = T_0 + (T_n - T_0)exp(-Bt) \qquad \text{Eq. 11}$$

Where $B = \frac{hs}{\sum m_i C_i}$, when the temperature is constant under the laser, $\frac{dT}{dt} = 0$ and $Q_{in} = Q_{ext}$, from which the photothermal conversion efficiency can be obtained as follows:

$$\eta = \frac{hs(T_n - T_0)}{I \cdot A_{abs}} \qquad \text{Eq. 12}$$

**Electronic Supplementary Material**

Supplementary material (size distribution analysis, SEM images of nanofilms, experimental measured and simulated UV-Vis-NIR spectrum and other information) is available in the online version of this article at http://dx.doi.org/XXXX/XXXX-***-****-* (automatically inserted by the publisher).

**Acknowledgement and Author Contribution**

Mr. Shenyi Ding synthesized and characterized the Au NS and Au NR nanoparticles, performed the photothermal experiment, taken photos of the photographs and experimental equipments (in **Figure 1** and **Figure 5**); Dr. Min Xi designed the experiment, performed the simulation, analyzed the data and wrote the manuscript; Dr. Cui Liu, Dr. Nian Li, Dr. Shudong Zhang, Dr. Jixiang Zhang and Dr. Zhenyang Wang provided the lab equipment and funding support.

The authors would acknowledge the finical support from the National Key Research and Development Project (2020YFA0210703), the Key Laboratory Functional Molecular Solids, Ministry of Education (FMS202002), the National Natural Science Foundation of China (Nos. U2032159, U2032158, and 62005292), the Major Scientific and Technological Special Project of Anhui Province (202103a05020192), the Key Research and Development Program of Anhui Province (202104a05020036), the special fund project for local science and technology development guided by

**Author Information**

Shenyi Ding – Email: dingshyi@163.com

Jixiang Zhang – Email: zhangjx@cqjtu.edu.cn

Cui Liu − Email: cliu0724@mail.ustc.edu.cn

Nian Li − Email: linian@issp.ac.cn

Shudong Zhang − Email: sdzhang@iim.ac.cn

Zhenyang Wang − Email: zywang@iim.ac.cn

Min Xi − Email: minxi@issp.ac.cn

**Reference**

(1) Cao, S.; Jiang, Q.; Wu, X.; Ghim, D.; Gholami Derami, H.; Chou, P. I.; Jun, Y. S.; Singamaneni, S. Advances in Solar Evaporator Materials for Freshwater Generation. *J. Mater. Chem. A*, **2019**, *7*, 24092–24123.

(2) Wu, S.-L.; Chen, H.; Wang, H.-L.; Chen, X.; Yang, H.-C.; Darling, S. B. Solar-Driven Evaporators for Water Treatment: Challenges and Opportunities. *Environ. Sci. Water Res. Technol.*, **2021**, *7*, 24–39.

(3) Zhang, C.; Liang, H. Q.; Xu, Z. K.; Wang, Z. Harnessing Solar-Driven Photothermal Effect toward the Water–Energy Nexus. *Adv. Sci.*, **2019**, *6*, 1900883.

(4) Eustis, S.; El-Sayed, M. Why Gold Nanoparticles Are More Precious than Pretty Gold: Noble Metal Surface Plasmon Resonance and Its Enhancement of the Radiative and Nonradiative Properties of Nanocrystals of Different Shapes. *Chem. Soc. Rev.*, **2006**, *35*, 209–217.

(5) Baffou, G.; Quidant, R. Thermo-Plasmonics: Using Metallic Nanostructures as Nano-Sources of Heat. *Laser Photonics Rev.*, **2013**, *7*, 171–187.

(6) Yeshchenko, O. A.; Bondarchuk, I. S.; Gurin, V. S.; Dmitruk, I. M.; Kotko, A. V. Temperature Dependence of the Surface Plasmon Resonance in Gold Nanoparticles. *Surf. Sci.*, **2013**, *608*, 275–281.

(7) Gao, M.; Peh, C. K.; Phan, H. T.; Zhu, L.; Ho, G. W. Solar Absorber Gel: Localized Macro-Nano Heat Channeling for Efficient Plasmonic Au Nanoflowers Photothermic Vaporization and Triboelectric Generation. *Adv. Energy Mater.*, **2018**, *8*, 1800710.

(8) Nazari, M.; Xi, M.; Lerch, S.; Alizadeh, M. H.; Ettinger, C.; Akiyama, H.; Gillespie, C.; Gummuluru, S.; Erramilli, S.; Reinhard, B. M. Plasmonic Enhancement of Selective Photonic Virus Inactivation. *Sci. Rep.*, **2017**, *7*, 11951.

(9) Nazari, M.; Xi, M.; Aronson, M.; McRae, O.; Hong, M. K.; Gummuluru, S.; Sgro, A. E.; Bird, J. C.; Ziegler, L. D.; Gillespie, C.; Souza, K.; Nguyen, N.; Smith, R. M.; Silva, E.; Miura, A.; Erramilli, S.; Reinhard, B. M. Plasmon-Enhanced Pan-Microbial Pathogen Inactivation in the Cavitation Regime: Selectivity without Targeting. *ACS Appl. Nano Mater.*, **2019**, *2*, 2548–2558.

(10) Jaque, D.; Martínez Maestro, L.; Del Rosal, B.; Haro-Gonzalez, P.; Benayas, A.; Plaza, J. L.; Martín Rodríguez, E.; García Solé, J. Nanoparticles for Photothermal Therapies. *Nanoscale*, **2014**, *6*, 9494–9530.

(11) Boyer, D.; Tamarat, P.; Maali, A.; Lounis, B.; Orrit, M. Photothermal Imaging of Nanometer-Sized Metal Particles among Scatterers. *Science*, **2002**, *297*, 1160–1163.

(12) Hong, Y.; Reinhard, B. M. Optoplasmonics : Basic Principles and Applications. *J. Opt.*, **2019**, *21*, 113001.

(13) Zhang, C.; Jia, F.; Li, Z.; Huang, X.; Lu, G. Plasmon-Generated Hot Holes for Chemical Reactions. *Nano Res.*, **2020**, *13*, 3183–3197.

(14) Zhang, Y.; He, S.; Guo, W.; Hu, Y.; Huang, J.; Mulcahy, J. R.; Wei, W. D. Surface-Plasmon-Driven Hot Electron Photochemistry. *Chem. Rev.*, **2018**, *118*, 2927–2954.

(15) Roller, E.-M.; Besteiro, L.; Pupp, C.; Khorashad, L.; Govorov, A.; Liedl, T. Hotspot-Mediated Non-Dissipative and Ultrafast Plasmon Passage. *Nat. Phys.*, **2017**, *13*, 761–765.

(16) Liang, J.; Liu, H.; Yu, J.; Zhou, L.; Zhu, J. Plasmon-Enhanced Solar Vapor Generation. *Nanophotonics*, **2019**, *8*, 771–786.

(17)	Zhuang, T. T.; Liu, Y.; Li, Y.; Zhao, Y.; Wu, L.; Jiang, J.; Yu, S. H. Integration of Semiconducting Sulfides for Full-Spectrum Solar Energy Absorption and Efficient Charge Separation. *Angew. Chemie - Int. Ed.*, **2016**, *55*, 6396–6400.

(18)	Li, Y.; Lin, C.; Wu, Z.; Chen, Z.; Chi, C.; Cao, F.; Mei, D.; Yan, H.; Tso, C. Y.; Chao, C. Y. H.; Huang, B. Solution-Processed All-Ceramic Plasmonic Metamaterials for Efficient Solar–Thermal Conversion over 100–727 °C. *Adv. Mater.*, **2021**, *33*, 2005074.

(19)	Xi, M.; Xu, L.; Li, N.; Zhang, S.; Wang, Z. Plasmonic $Cu_{27}S_{24}$ Nanocages for Novel Solar Photothermal Nanoink and Nanofilm. *Nano Res.*, **2021**.

(20)	Xi, M.; Reinhard, B. M. Localized Surface Plasmon Coupling between Mid-IR-Resonant ITO Nanocrystals. *J. Phys. Chem. C*, **2018**, *122*, 5698–5704.

(21)	Zhuo, X.; Yip, H.-K.; Ruan, Q.; Zhang, T.; Zhu, X.; Wang, J.; Lin, H.; Xu, J.; Yang, Z. Broadside Nanoantennas Made of Single Silver. *ACS Nano*, **2018**, *12*, 1720–1731.

(22)	Zhuo, X. Colour Routing with Single Silver Nanorods. *Light Sci. Appl.*, **2019**, *8*, 1–11.

(23)	Xi, M.; Reinhard, B. M. Evolution of Near- and Far-Field Optical Properties of Au Bipyramids upon Epitaxial Deposition of Ag. *Nanoscale*, **2020**, *12*, 5402–5411.

(24)	Liz-Marzan, L. Tailoring Surface Plasmons through the Morphology and Assembly of Metal Nanoparticles. *Langmuir*, **2006**, No. 22, 32–41.

(25)	Burrows, N.; Lin, W.; Hinman, J.; Dennison, J.; Vartanian, A.; Abadeer, N.; Grzincic, E.; Jacob, L.; Li, J.; Murphy, C. Surface Chemistry of Gold Nanorods. *Langmuir*, **2016**, *32*, 9905–9921.

(26)	Wu, L.; Reinhard, B. M. Probing Subdiffraction Limit Separations with Plasmon Coupling Microscopy: Concepts and Applications. *Chem. Soc. Rev.*, **2014**, *43*, 3884–3897.

(27)	Nordlander, P.; Oubre, C.; Prodan, E.; Li, K.; Stockman, M. I. Plasmon Hybridization in Nanoparticle Dimers. *Nano Lett.*, **2004**, *4*, 899–903.

(28)	Jain, P. K.; Huang, W.; El-Sayed, M. A. On the Universal Scaling Behavior of the Distance Decay of Plasmon Coupling in Metal Nanoparticle Pairs: A Plasmon Ruler Equation. *Nano Lett.*, **2007**, *7*, 2080–2088.

(29)	Harris, N.; Arnold, M. D.; Blaber, M. G.; Ford, M. J. Plasmonic Resonances of Closely Coupled Gold Nanosphere Chains. *J. Phys. Chem. C*, **2009**, *113*, 2784–2791.

(30) Chen, T.; Pourmand, M.; Feizpour, A.; Cushman, B.; Reinhard, B. M. Tailoring Plasmon Coupling in Self-Assembled One-Dimensional Au Nanoparticle Chains through Simultaneous Control of Size and Gap Separation. *J. Phys. Chem. Lett.*, **2013**, *4*, 2147–2152.

(31) Yan, B.; Thubagere, A.; Premasiri, W. R.; Ziegler, L.; Negro, L. D.; Reinhard, B. Engineered SERS Substrates with Multiscale Signal Enhancement: Nanoparticle Cluster Arrays. *ACS Nano*, **2009**, *3*, 1190–1202.

(32) Hong, Y.; Ahn, W.; Boriskina, S.; Zhao, X.; Reinhard, B. Directed Assembly of Optoplasmonic Hybrid Materials with Tunable Photonic–Plasmonic Properties. *J. Phys. Chem. Lett.*, **2015**, *6*, 2056–2064.

(33) Chen, M.; He, Y.; Zhu, J. Quantifying and Comparing the Near-Field Enhancement, Photothermal Conversion, and Local Heating Performance of Plasmonic $SiO_2$@Au Core-Shell Nanoparticles. *Plasmonics*, **2019**, *14*, 1019–1027.

(34) Bae, K.; Kang, G.; Cho, S. K.; Park, W.; Kim, K.; Padilla, W. J. Flexible Thin-Film Black Gold Membranes with Ultrabroadband Plasmonic Nanofocusing for Efficient Solar Vapour Generation. *Nat. Commun.*, **2015**, *6*.

(35) Chen, M.; Wu, Y.; Song, W.; Mo, Y.; Lin, X.; He, Q.; Guo, B. Plasmonic Nanoparticle-Embedded Poly(p-Phenylene Benzobisoxazole) Nanofibrous Composite Films for Solar Steam Generation. *Nanoscale*, **2018**, *10*, 6186–6193.

(36) Malachosky, E.; Guyot-Sionnest, P. Gold Bipyramid Nanoparticle Dimers. *J. Phys. Chem. C*, **2014**, *118*, 6405–6412.

(37) Slaughter, L.; Wu, Y.; Willingham, B.; Nordlander, P.; Link, S. Effects of Symmetry Breaking and Conductive Contact on the Plasmon Coupling in Gold Nanorod Dimers. *ACS Nano*, **2010**, *4*, 4657–4666.

(38) Willets, K.; Van Duyne, R. Localized Surface Plasmon Resonance Spectroscopy and Sensing. *Annu. Rev. Phys. Chem.*, **2007**, *58*, 267–297.

(39) Brown, L.; Yang, X.; Zhao, K.; Zheng, B.; Nordlander, P.; Halas, N. Fan-Shaped Gold Nanoantennas Above Reflective Substrates for Surface-Enhanced Infrared Absorption (SEIRA). *Nano Lett.*, **2015**, *15*, 1272–1280.

(40) Bauch, M.; Toma, K.; Toma, M.; Zhang, Q.; Dostalek, J. Plasmon-Enhanced Fluorescence Biosensors: A Review. *Plasmonics*, **2014**, *9*, 781–799.

(41)    Li, J. F.; Li, C. Y.; Aroca, R. F. Plasmon-Enhanced Fluorescence Spectroscopy. *Chem. Soc. Rev.*, **2017**, *46*, 3962–3979.

(42)    Felix-Rendon, U.; Berini, P.; De Leon, I. Ultrasensitive Nanoplasmonic Biosensor Based on Interferometric Excitation of Multipolar Plasmonic Modes. *Opt. Express*, **2021**, *29*, 17365.

(43)    Aubry, A.; Lei, D.; Maier, S.; Pendry, J. Interaction Between Plasmonic Nanoparticles Revisited with Transformation Optics. *Phys. Rev. Lett.*, **2010**, *105*, 233901.

(44)    Liu, N.; Weiss, T.; Mesch, M.; Langguth, L.; Eigenthaler, U.; Hirscher, M.; Sönnichsen, C.; Giessen, H. Planar Metamaterial Analogue of Electromagnetically Induced Transparency for Plasmonic Sensing. *Nano Lett.*, **2010**, *10*, 1103–1107.

(45)    Yang, Z.; Zhang, Z.; Zhang, L.; Li, Q.; Hao, Z.; Wang, Q. Fano Resonances in Dipole-Quadrupole Plasmon Coupling Nanorod Dimers. *Opt. Lett.*, **2011**, *36*, 1542–1544.

(46)    Zhang, S.; Geryak, R.; Geldmeier, J.; Kim, S.; Tsukruk, V. V. Synthesis, Assembly, and Applications of Hybrid Nanostructures for Biosensing. *Chem. Rev.*, **2017**, *117*, 12942–13038.

(47)    Liu, R.; Jiang, L.; Liu, G.; Chen, G.; Li, J.; Liu, J.; Wang, X. Optical Response Properties of Stable and Controllable Au Nanorod. *J. Phys. Chem. C*, **2019**, *123*, 13892–13899.

(48)    Li, Q.; Zhuo, X.; Li, S.; Ruan, Q.; Xu, Q. H.; Wang, J. Production of Monodisperse Gold Nanobipyramids with Number Percentages Approaching 100% and Evaluation of Their Plasmonic Properties. *Adv. Opt. Mater.*, **2015**, *3*, 801–812.

(49)    He, D.; Hu, B.; Yao, Q.; Wang, K.; Yu, S. Large-Scale Synthesis of Flexible Free-Sensitivity: Electrospun PVA Nanofibers of Silver Nanoparticles. *ACS Nano*, **2009**, *3*, 3993–4002.

(50)    Zhu, X. Z.; Zhuo, X.; Li, Q.; Yang, Z.; Wang, J. F. Gold Nanobipyramid-Supported Silver Nanostructures with Narrow Plasmon Linewidths and Improved Chemical Stability. *Adv. Funct. Mater.*, **2016**, *26*, 341–352.

(51)    Vigderman, L.; Zubarev, E. High-Yield Synthesis of Gold Nanorods with Longitudinal SPR Peak Greater than 1200 Nm Using Hydroquinone as a Reducing Agent. *Chem. Mater.*, **2013**, *25*, 1450–1457.

(52)    Ding, W.; Zhang, P.; Li, Y.; Xia, H.; Wang, D.; Tao, X. Effect of Latent Heat in Boiling Water on the Synthesis of Gold Nanoparticles of Different Sizes by Using the Turkevich Method. *ChemPhysChem*, **2015**, *16*, 447–454.

# Supporting Information

# Investigation of plasmonic enhanced solar photothermal effect of Au NR@PVDF micro/nano-film

Shenyi Ding[a,b], Jixiang Zhang*[a,b], Cui Liu[b], Nian Li[b], Shudong Zhang[b], Zhenyang Wang*[b], Min Xi*[b,c]

[a] School of Mechatronics and Vehicle Engineering, Chongqing Jiaotong University, Chongqing, 400074, P. R. China

[b] Institute of Solid State Physics and Key Laboratory of Photovoltaic and Energy Conservation Materials, Hefei Institutes of Physical Science, Chinese Academy of Sciences, Hefei, Anhui 230031, P. R. China

[c] The Key Laboratory Functional Molecular Solids Ministry of Education, Anhui Normal University, Wuhu, Anhui, 241002, P. R. China

Corresponding authors:

Jixiang Zhang − Email: zhangjx@cqjtu.edu.cn

Zhenyang Wang − Email: zywang@iim.ac.cn

Min Xi − Email: minxi@issp.ac.cn

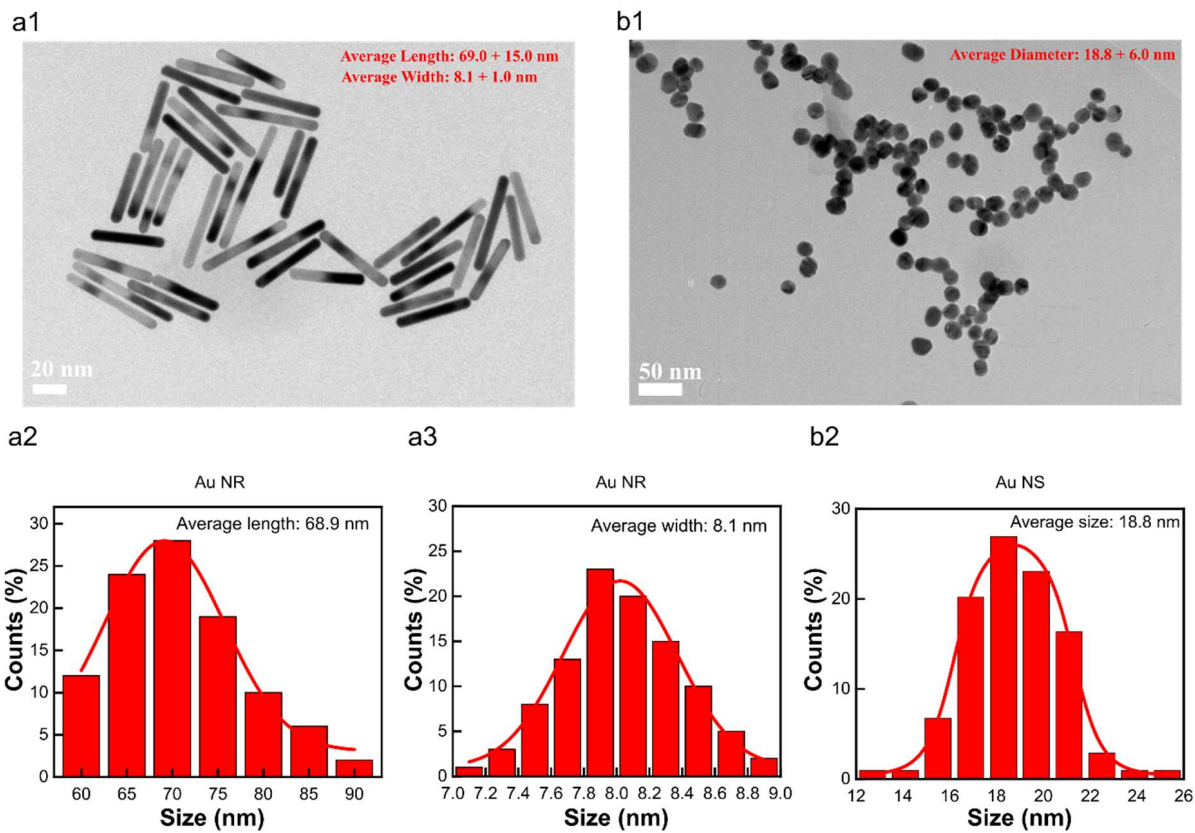

**Figure S1** TEM of Au NR **(a1)** and Au NS **(b1)**; Statistical count of Au NR (**(a2)** and **(a3)**) and Au NS (**b2**) size distributions.

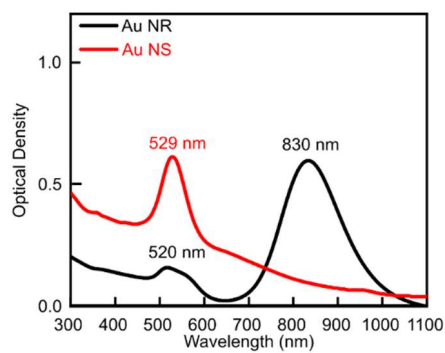

**Figure S2** UV-Vis-NIR spectrum of Au NR (black line) and Au NS (red line) aqueous solution, respectively.

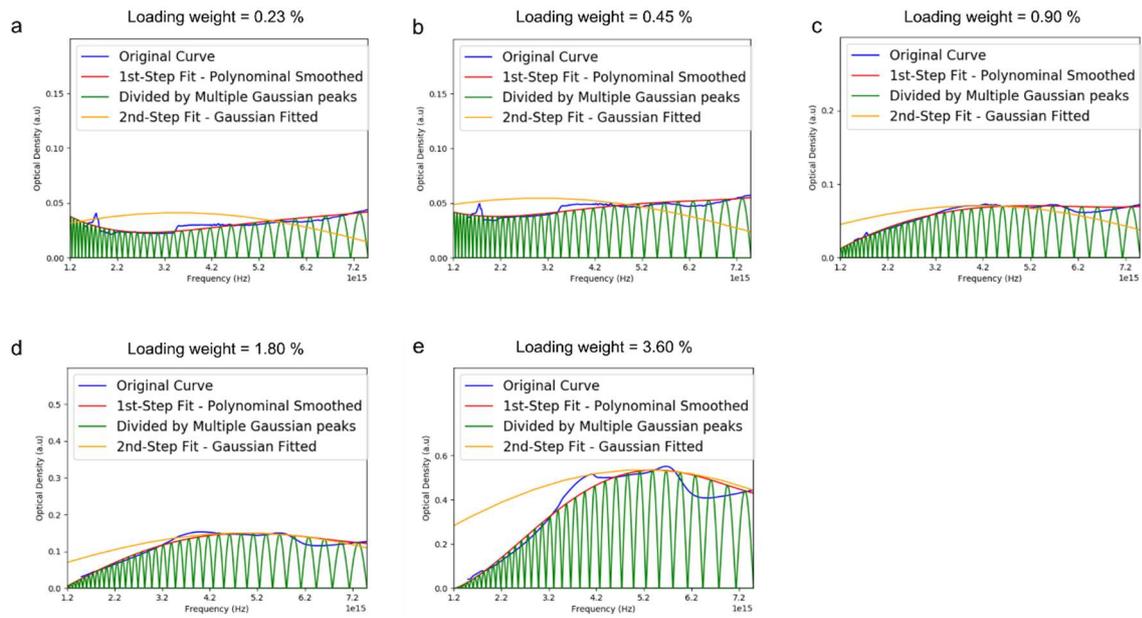

**Figure S3** UV-Vis-NIR spectrum analysis of Au NR/PVDF nanofilms with different Au NR loading concentrations from 0.23 wt% in **Figure 3a** to 3.60 wt% in **Figure 3e**.

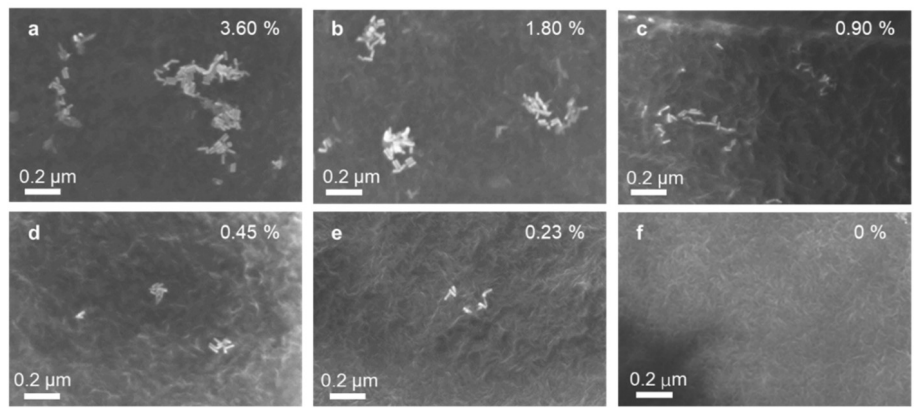

**Figure S4** Representative SEM images of Au NR/PVDF nanofilm with Au loaded weight concentrations varied from 3.60 % (in **Figure S4a**) to 0.00 % (in **Figure S4f**).

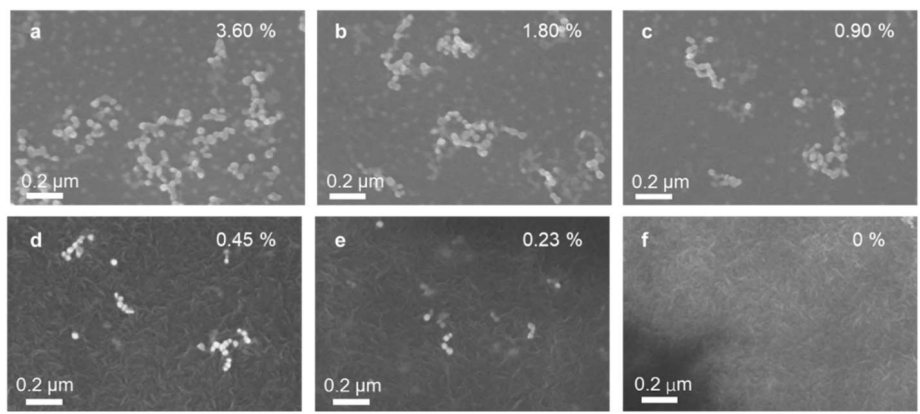

**Figure S5** Representative SEM images of Au NS/PVDF nanofilm with Au loaded weight concentrations varied from 3.60 % (in **Figure S5a**) to 0.00 % (in **Figure S5f**).

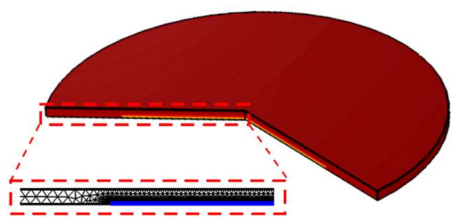

**Figure S6** Simulation mesh plot.

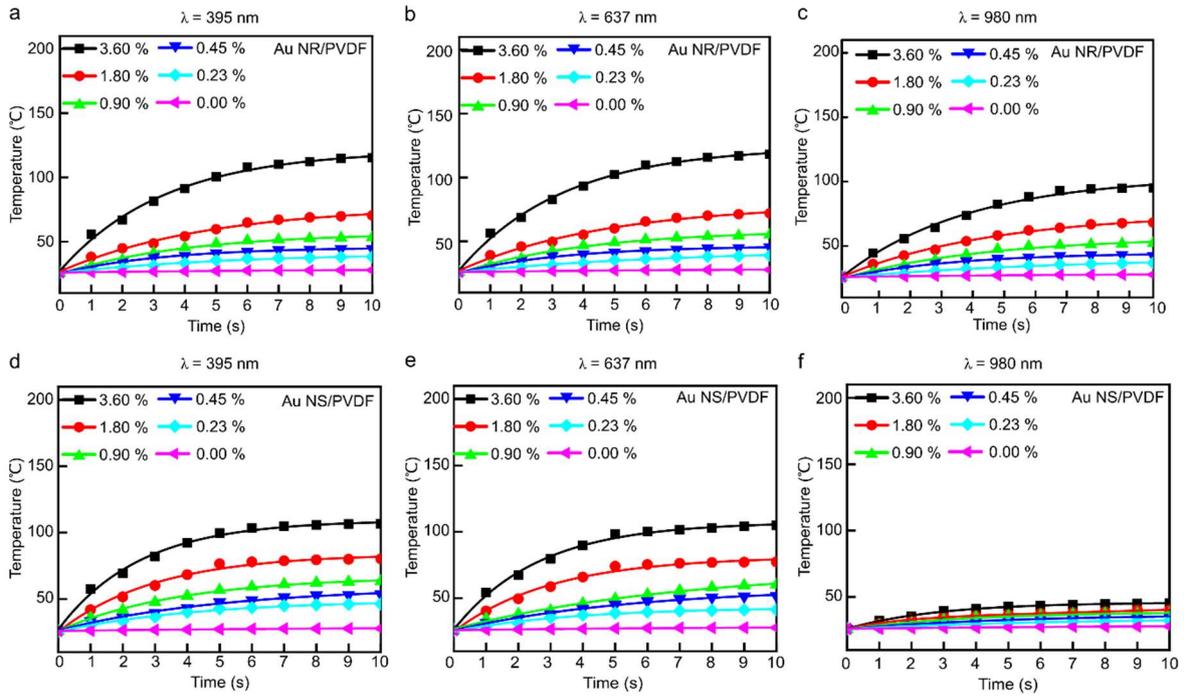

**Figure S7** Experimental measured temperature variation of Au/PVDF films under different laser irradiation. **(a) - (c)** Temperature variation of Au NR/PVDF films irradiated by 395 nm, 637 nm and 980 nm laser; **(d) - (f)** Temperature variation of Au NS/PVDF films irradiated by 395 nm, 637 nm and 980 nm laser.

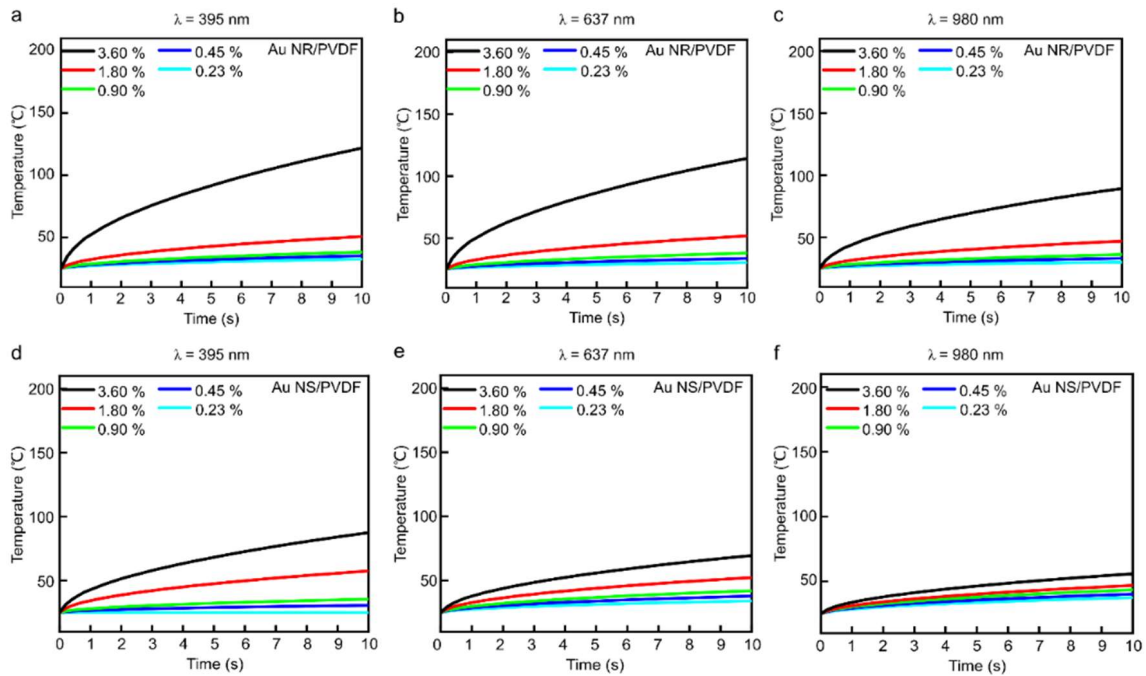

**Figure S8** Simulated temperature variation of Au/PVDF films under different laser irradiation. **(a) - (c)** Temperature variation of Au NR/PVDF films irradiated by 395 nm, 637 nm and 980 nm laser; **(d) - (f)** Temperature variation of Au NS/PVDF films irradiated by 395 nm, 637 nm and 980 nm laser.

**Table S1** Solar photothermal conversion efficiency

| Weight | η (Au NR) | η (Au NS) |
|---|---|---|
| 0.23 % | 1.3 % | 1.1 % |
| 0.45 % | 3.4 % | 3.2 % |
| 0.90 % | 6.7 % | 6.4 % |
| 1.80 % | 9.4 % | 9.1 % |
| 3.60 % | 12.1 % | 11.8 % |

**Table S2** Monolaser photothermal conversion efficiency

| Monolaser | η (Au NR) | η (Au NS) |
|---|---|---|
| 395 nm | 85.4 % | 76.9 % |
| 637 nm | 88.3 % | 75.9 % |
| 980 nm | 79.5 % | 70.4 % |